\documentclass[twocolumn,tighten]{aastex631}

\usepackage[all]{hypcap} %Figure refs breaks deluxetables; use %\capstartfalse %\capstarttrue to fix it

\usepackage{url}
\usepackage{paralist}
\usepackage{float}

\usepackage{subfigure}

\shorttitle{Cadence impacts on reliable classification of variable stars}
\shortauthors{Hernitschek et al.}
\graphicspath{{./}{figures/}}
\begin{document}

\title{The Impact of Observing Strategy on Reliable Classification of Standard Candle Stars: Detection of Amplitude, Period, and Phase Modulation (Blazhko Effect) of RR Lyrae Stars with LSST}

\author{Nina Hernitschek}
\altaffiliation{nina.hernitschek@vanderbilt.edu}
\affiliation{Department of Physics \& Astronomy, Vanderbilt University, Nashville, TN 37212, USA}
\affiliation{Data Science Institute, Vanderbilt University, Nashville, TN 37212, USA}

\author{Keivan G. Stassun}
\affiliation{Department of Physics \& Astronomy, Vanderbilt University, Nashville, TN 37212, USA}

%  250 words maximum; this abstract has exactly 250 words
\begin{abstract}
The Vera C. Rubin Observatory will carry out its Legacy Survey of Space and Time (LSST) 
with a single-exposure depth of $r{\sim}24.7$ and an anticipated baseline of 10 years, allowing to access the Milky Way's old halo not only deeper, but also with a longer baseline and better cadence than e.g. PS1 3$\pi$ \citep{Chambers2016}. This will make LSST ideal to study populations of variable stars such as RR Lyrae stars (RRL). Here, we address the question of observing strategy optimization of LSST, as survey footprint definition, single visit exposure time as well as the cadence of repeat visits in different filters are yet to be finalized. 
We present metrics used to assess the impact of different observing strategies on the reliable detectability and classification of standard-candle variable stars, including detection of amplitude period, phase modulation effects of RR Lyrae stars, the so-called Blazhko effect \citep{Blazhko1907, Kollath2011}, by evaluating metrics for simulated potential survey designs.
So far, due to depth and cadence of typical all-sky surveys, it was nearly impossible to study this effect on a larger sample. All-sky surveys with relatively few observations over a moderately long baseline allow only for fitting phase-folded RRL light curves, thus integrating over the complete survey length and hiding any information regarding possible period or phase modulation during the survey. On the other hand, surveys with a cadence to detect slightly changing light curves usually have a relatively small footprint. LSST's survey strategy, however, will allow for studying variable stars in a way that makes population studies possible.
\end{abstract}

\section{Introduction}
\label{sec:Introduction}

The upcoming Legacy Survey of Space and Time (LSST) is designed to meet a broad range of science goals with a single 10-year, time-domain survey.
LSST's four key science pillars are: Constraining Dark Energy \& Dark Matter, Taking an Inventory of the Solar System, Exploring the Transient Optical Sky and
Mapping the Milky Way.

With a single-exposure depth of $r{\sim}24.7$, its long temporal baseline, and its its ability to cover a third of the sky each night, detecting billions of stars and galaxies, and millions of transients and variable objects, this photometric survey will be well equipped for these science cases. Especially its ability to carry out repeated observations over a wide sky area, will deliver unprecedented advances in a diverse set of science cases \citep[][hereafter The LSST Science Book]{Abell2009}.

As members of the LSSTC Transients and Variable Stars (TVS) Classification group, we here take a closer look at how the cadence choice will influence LSST's ability to detect characteristic modulations of amplitude, period and phase of RR Lyrae \citep[the so-called Blazhko effect,][] {Blazhko1907}, which falls into the science goals of mapping the Milky Way and exploring the transient and variable optical sky.

So far, due to depth and cadence of typical all-sky surveys, it was nearly impossible to study this effect on a larger sample. Surveys such as PS1 3$\pi$ \citep{Chambers2016} with relatively few observations over a moderately long baseline allow only for fitting the period and phase of RR Lyrae stars while integrating over the complete survey length, thus not giving any information regarding whether the period and/or phase of the light curve might have changed during the survey \citep{Hernitschek2016,Sesar2017a}. On the other hand, surveys specialized for detecting slightly changing light curves due to very finely sampled cadence usually have either an input catalog \citep[such as TESS does, see][]{Ricker2015}, or a relatively small footprint, which doesn't allow for reliable statistics over large samples. LSST's ability to cover a wide area of the sky while making deep, rapid observations without using an input catalog, however, will allow for studying variable stars in the Milky Way's old halo in a way that makes population studies possible.

Around 80\% of the survey's observing time will be dedicated to the main or Wide Fast Deep (WFD) survey, covering at least 18,000 deg$^2$. In addition to that, LSST will carry out ``mini-surveys'' (for instance, a dedicated Galactic plane survey), ``deep drilling fields'' (DDFs) and potentially ``targets of opportunity'' (ToOs).

Covering this vast amount of science goals makes the choice of an observing strategy an important, but also difficult, problem. We emphasize here the community-driven approach for LSST survey strategy optimization, which started with an observing strategy white paper \citep{LSST2017} and led to the LSST Project
Science Team and the LSST Science Advisory Committee releasing a call for community white papers proposing
observing strategies for the LSST WFD survey, as well as the DDFs and mini-surveys \citep{Ivezic2018}. Survey strategies building on those white papers are discussed in \cite{Jones2021}. 
Finally, the LSST Survey Cadence Optimization Committee\footnote{\url{https://www.lsst.org/content/charge\%2Dsurvey-cadence-optimization-committee-scoc}} (SCOC) released a series of top-level survey strategy questions\footnote{\url{https://docushare.lsst.org/docushare/dsweb/Get/Document-36755}} which can be analyzed using the simulations of \cite{Jones2021}, leading to ``cadence notes''.

In response to this, a white paper on a mini-survey of the northern sky to Dec $<$+30 \citep{Capak2018}, as well as a paper on
the impact of LSST template acquisition strategies on early science for transients and variable stars \citep{Hambleton2020} were created, on which the first author of this paper is a co-author. Recently, we submitted a cadence note\footnote{\url{https://docushare.lsst.org/docushare/dsweb/Get/Document-37673}} which was the precursor of this paper.

In this paper, in order to support decisions on the survey strategy, we develop a quantitative metric, as well as evaluate a number of simulated LSST observing strategies on this metric regarding their ability to detect the characteristic RR Lyrae light-curve modulations of the Blazhko effect.
This paper is part of the ApJS Rubin Cadence Focus Issues, for which an opening paper is available \citep{Bianco2021}.

The paper is structured as follows:
Section \ref{sec:ScienceCases} describes the science cases we investigate. Section \ref{sec:StudyParametersandMethods} gives a brief overview of anticipated observing strategies for LSST, the sets of simulations of different observing strategies, and how we access those data.
In Section \ref{sec:Results}, we present metrics that are capable of assessing the impact of different observing strategies on those science cases, as well as how we analyze the results we get from running the metric we developed on different cadence simulations. In Section \ref{sec:OptimizationofFilterandCadenceAllocation}, we discuss our results drawn from the analysis of the metric on the simulated observing strategy, as well as highlight our recommendations for the observing strategy choice regarding our science goals. Finally, in Section \ref{sec:Conclusions}, we summarize our conclusions drawn from the analysis of the metric on the simulated observing strategy and our recommendations for the observing strategy choice regarding our science goals.
 
\section{Science Cases} 
\label{sec:ScienceCases}

Pulsating stars of the RR Lyrae type have been studied for over a century now, as they play an important role in distance estimation as well as in studying the old halo content, such as globular clusters and stellar streams, of our Milky Way. RR Lyrae stars serve as relatively easily detectable standard candles, enabling the calculation of distances from PLZ (period, luminosity, metallicity) relationships.
Depending on their light-curve characteristics, RR Lyrae stars are divided into subclasses. The RRab stars, pulsating in the fundamental radial mode, show high-amplitude, skewed non-sinusoidal light curves. RRc stars, in contrast, pulsate in the radial first overtone mode and show a nearly sinusiodal light curve with smaller amplitudes than RRab stars. RRd stars pulsate in both modes simultaneously.

The Blazhko effect, first observed by S. Blazhko in 1907 in the star RW Draconis \citep{Blazhko1907}, is a modulation of period, phase and amplitude in RR Lyrae stars. The typical pulsation period of an RR Lyrae star is 0.2 - 1.1 days with a 0.5 - 1 mag amplitude, whereas the modulation due to the Blazhko effect is on timescales from weeks to months, and its amplitude is a few tenths of magnitudes smaller. The Blazhko effect is observed for about 20 - 30\% of the RRab stars \citep{Szeidl1988, MoskalikPoretti2003} and for less than 5\% of the RRc stars.

The physics behind the Blazhko effect is still under discussion. Among the three primary hypotheses, the first (resonance model)
sees the cause of the modulation in a nonlinear resonance among the fundamental or the first-overtone pulsation mode and a higher mode. The second (magnetic model) sees the cause of the modulation in the magnetic field being inclined to the rotational axis of the star, thus deforming the main radial mode. The third model assumes cycles occurring in the convection as cause for the modulations.
The cadence, depth and large footprint of LSST will allow us to address questions such as how frequent the Blazhko effect is, and in addition will allow us to carry out population studies on stars showing these modulation effects. Such studies then can shed more light on the question of underlying physical processes driving the Blazhko effect.
Also, as using RR Lyrae stars as standard candles, determining correct periods is crucial for calculating distances. Better understanding the impact of the Blazhko effect, and taking it into account when calculating distances, will improve our ability to study the halo population of our Milky Way.

\section{Study Parameters and Methods} 
\label{sec:StudyParametersandMethods}

In this section, we give a brief overview of LSST's anticipated observing strategies under discussion for the main WFD survey, as well as the simulations used for evaluating survey strategies within the Metrics Analysis Framework.

\subsection{LSST Observing Strategy}
\label{sec:LSSTObservingStrategy}

Currently under construction, the Rubin Observatory located at Cerro Pach\'{o}n in Northern Chile, and will undertake the Legacy
Survey of Space and Time (LSST), a 10-year survey expected to start in 2023.
With an 8.4 m (6.7 m effective) diameter primary mirror and a 9.6 deg$^2$ field of view, the telescope will reach a 5$\sigma$ single-exposure depth (over 30 s) in ugrizy of
23.9, 25.0, 24.7, 24.0, 23.3, 22.1 mag. The co-added survey depth will be approximately 26.1, 27.4, 27.5, 26.8, 26.1, 24.9 mag.
An alternative exposure strategy, instead of taking 30 s exposures, is taking two successive exposures of 15 s. The latter will help to 
mitigate cosmic-ray and satellite trail artifacts. The decision between these exposure strategies will not be made until the commissioning phase of the LSST survey.

Detailed technical specifications of the Rubin Observatory and LSST can be found in the LSST Science Requirements Document \citep[][hereafter SRD]{Ivezic2013}.

The observing strategy of LSST is impacted by several factors, and its optimization to take into account the survey’s wide range of science drivers is especially challenging. 
The LSST SRD lists the following top specifications for the survey:
\begin{itemize}
\item The sky area uniformly covered by the main survey (WFD) will include at least 15,000 deg$^2$, and 18,000 deg$^2$ by design.
\item The sum of the median number of visits in each band across the sky will be at least 750, with a design goal of 825.
\item At least 1,000 deg$^2$ of the sky, with a design goal of 2,000 deg$^2$, will have multiple observations separated by
nearly uniformly sampled time scales ranging from 40 sec to 30 min.
\end{itemize}

The document also emphasizes the influence of observing conditions such as seeing, sky brightness (affected by lunar phase and time of night), transparency, and airmass. The algorithm of the observation scheduler will balance observing goals to maintain a coverage as uniform as possible in time and in location on the sky.

Given the specifications described above, a baseline observing strategy was developed \citep{Jones2021}.

The strategy strategy outlined here is considered the current nominal observing strategy plan pending further modifications, as outlined in \citep{Jones2021}:
\begin{itemize}
\item Visits are 1 $\times$ 30 s long. According to the baseline simulation, this results in ${\sim} 2.2 \times 10^6$ visits over the anticipated baseline of 10 years.
\item To assist with asteroid identification, pairs of visits are carried out. These pairs of visits are scheduled for approximately 22 min separation and are in two filters as follows: $u - g$, $u - r$, $g - r$, $r - i$, $i - z$, $z - y$ or $y - y$. 
\item The survey footprint is composed of the main Wide-Fast-Deep survey (WFD, spanning 18,000 deg$^2$ from -62 to +2 degrees excluding the Galactic equator), the North Ecliptic Spur (NES), the Galactic Plane (GP) and South Celestial Pole (SCP). The baseline footprint includes WFD, NES
($griz$ only), GP, and SCP. WFD is ${\sim}82$\% of the total survey time.
\item Deep-drilling fields (DDF) with deeper coverage and a more frequent temporal sampling. The baseline strategy includes five DDFs, 
one of them being composed of two pointings covering the Euclid Deep Field-South (EDF-S). DDF make up 5\% of the total survey time.
\item Visits will be distributed between the filters as follows: 6\% in $u$, 9\% in $g$, 22\% in $r$, 22\% in $i$, 20\% in $z$, and 21\% in $y$.
\item As the filter exchange system can hold five out of the six filters at the same time, the decision which five filters will actually be installed in the filter exchange system is made as follows: If at the start of the night
the moon is 40\% illuminated or more (corresponding to approximately full moon $\pm$ 6 nights), the $z$-band filter is installed; otherwise the $u$-band filter is installed.
\item The camera is rotationally dithered nightly between -80 and 80 degrees. 
Rotating the camera will cancel field rotation during an exposure; after that, the camera reverts back to the chosen rotation angle for the next exposure. At the beginning of each night, the rotation is randomly selected.
\item Additional twilight observations in the morning and evening twilight carried out in $rizy$ during nights where the twilight survey is active, and are chosen by an algorithm that maximizes the reward function at each observation. Typically about 440 1 s visits per night are carried out in the twilight survey. $ug$ filters are not part of the twilight survey.
\item The baseline strategy observes the entire footprint each observing season (``non-rolling cadence''). Under discussion is also a rolling
cadence that would prioritize sections of the footprint at different times (e.g., observing half the footprint for one year
and changing to the other half the next). This would improve the cadence in that section.
\end{itemize}

\subsection{Survey and Observing Strategy Simulations}
\label{sec:SurveyandObservingStrategySimulations}

In order to allow for evaluating metrics for potential survey designs, survey simulations have been created.

LSST Operations Simulation \citep[Opsim,][]{Naghib2019, Delgado2016, Delgado2014} generates simulated pointing histories under realistic conditions by using the default scheduler for LSST, the Feature-Based Scheduler \citep[FBS,][]{Naghib2019}. FBS utilizes a modified Markov Decision Process to decide the next observing direction and filter selection, allowing a flexible
approach to scheduling, including the ability to compute a reward function throughout a night.

New versions of simulated strategies are released regularly with improvements to the scheduler, weather simulation and changes to the baseline strategy. In this paper, we focus on the latest version, LSST OpSim FBS 1.7.
In Table \ref{tab:simulationfamilies}, we give a summary of the different simulations (the so-called survey strategy simulation families) in LSST OpSim FSB 1.7.

A detailed description LSST OpSim FSB 1.7, including recent survey strategy changes and summary statistics, can be found in the in official Rubin Observatory online resources, i.e. the OpSim1.7 Summary Information Document\footnote{\url{https://github.com/lsst-pst/survey_strategy/blob/master/fbs_1.7/SummaryInfo.ipynb}} and the Survey Simulations v1.7 release (January 2021)\footnote{\url{https://community.lsst.org/t/survey\%2Dsimulations-v1-7-release-january-2021/4660}}.
The simulations can be downloaded at \url{http://astro-lsst-01.astro.washington.edu:8081/}.

\clearpage

\begin{longtable*}{p{1.8cm}p{5.5cm}p{8cm}} 
\caption{Description of LSST OpSim FSB 1.7 release simulation families, including a brief explanation of the family.}\label{tab:simulationfamilies}\\
 \hline
 simulation family & simulation runs & Description \\
 \hline
baseline & \texttt{baseline\_nexp1\_v1.7\_10yrs.db}, \texttt{baseline\_nexp2\_v1.7\_10yrs.db} & Simulations of the baseline strategy, with 1$\times$30 s and 2$\times$15 s visits, respectively. The baseline survey footprint includes the main Wide-Fast-Deep (WFD) survey, several mini-surveys and the so-called Deep Drilling Fields (DDF).
All other runs in LSST OpSim FSB 1.7 have 2$\times$15 s visits and thus should be compared to \texttt{baseline\_nexp2\_v1.7\_10yrs.db}. \\
footprint\_tune & \texttt{footprint\_\{i\}\_v1.7\_10yrs},\newline \texttt{i = \{0,...,8\}} & Test of varying survey footprints based on increasing the low-extinction area of the WFD; coverage of the bulge and outer Milky Way disk as part of the WFD area is also tested.\\
rolling & \texttt{rolling\_\{i\}scale\{j\}\_nslice\{k\}\_}\newline\texttt{v1.7\_10yrs}, \newline \texttt{i = \{,nm\_\}},\newline  \texttt{j = \{0.2, 0.4, 0.6, 0.8, 0.9, 1.0\}},  \texttt{k=\{2,3\}} & 
Rolling cadence involves that some parts of the sky receive a higher number of visits during an `on' season, followed by a lower number of visits during an 'off' season. Test of different rolling cadence scenarios by dividing the sky in half (nslice2) or thirds (nslice3), varying the weight of the rolling from 20\% (scale0.2) to 100\% (scale1.0). In addition, with \texttt{nm} rolling simulations are tested which do not include the daily north/south modulation.\\ 
ddf\_dithers & \texttt{ddf\_dither\{i\}\_v1.7\_10yrs}, \newline \texttt{i = \{0, 0.05, 0.10, 0.30, 0.70, 1.00, 1.50, 2.00\}} & Test of the size of the translational dither offsets in the DDF, varying from 0 to 2 degrees. Smaller dithers will help the overall depth and uniformity, but larger dithers may be needed for calibration.\\
euclid\_dithers & \texttt{euclid\_dither\{i\}\_v1.7\_10yrs}, \newline \texttt{i = \{1,...,5\}} & Tests of varying translational dither offsets for the Euclid DDF footprint. The simulations vary the spacial dither both towards and perpendicular to the second pointing.\\
pair\_times & \texttt{pair\_times\_\{i\}\_v1.7\_10yrs}, \newline \texttt{i= \{11, 22, 33, 44, 55\} } & Tests of varying pair time from 11 to 55 minutes. In the baseline simulation, observations are typically taken in pairs separated by ${\sim}$22 minutes. \\  
twilight\_pairs & \texttt{twi\_pairs\_v1.7\_10yrs, twi\_pairs\_mixed\_v1.7\_10yrs, twi\_pairs\_repeat\_v1.7\_10yrs, twi\_pairs\_mixed\_repeat\_v1.7\_10yrs} & Test of pair twilight visits rather than single visits as in the baseline strategy.
The baseline simulations to not attempt to pair twilight visits. Depending on the simulations, twilight observations are paired in either the same filter, or a different filter (mixed). The twilight observations are also set to attempt to re-observe areas of the sky that have already been observed in the night (repeat). The simulation are as follows: Twilight pairs same filter, twilight pairs in mixed filters, twilight pairs in in the same filter with repeated area, twilight pairs in mixed filters with repeated area\\
twi\_neo & \texttt{twi\_neo\_pattern\{i\}\_v1.7\_10yrs},  \texttt{i=\{1,...,7\}} & Test of the impact of adding a twilight NEO survey which includes a large number of 1s observations. These visits are acquired in both morning and evening twilight, in sets of triplets separated by about 3 minutes.\\
u\_long & \texttt{u\_long\_ms\_\{i\}\_v1.7\_10yrs}, \newline \texttt{i=\{30, 40, 50, 60\}} & Test of different $u$ band exposure times. Observations in the $u$ filter are taken as single exposures. The number of $u$ band visits is left unchanged, resulting in a shift of visits from other filters to compensate for the increase in time. (Exception: DDF $u$-band observations are the default 2$\times$15s exposures).\\ 
filter\_cadence & \texttt{cadence\_drive\_gl\{i\}\_\{j\}v1.7\_10yrs},  \texttt{i=\{30,100,200\}},  \texttt{j=\{, gcb\} } & Test of impact of reducing gaps between $g$ band visits over the month. Different limits on how many $g$-band observations are taken. Long gaps in the $g$-band exposures are avoided by fill-in visits. \\
new\_rolling & \texttt{rolling\_nm\_scale0.90\_nslice2\_}\newline\texttt{fpw0.9\_nrw1.0v1.7\_10yrs},  \texttt{rolling\_nm\_scale0.90\_nslice3\_}\newline\texttt{fpw0.9\_nrw1.0v1.7\_10yrs},  \texttt{full\_disk\_v1.7\_10yrs}, 
 \texttt{full\_disk\_scale0.90\_nslice2\_}\newline\texttt{fpw0.9\_nrw1.0v1.7\_10yrs},  \texttt{full\_disk\_scale0.90\_nslice3\_}\newline\texttt{fpw0.9\_nrw1.0v1.7\_10yrs},  \texttt{footprint\_6\_v1.7.1\_10yrs},  \texttt{bulge\_roll\_scale0.90\_nslice2\_}\newline\texttt{fpw0.9\_nrw1.0v1.7\_10yrs},  \texttt{bulge\_roll\_scale0.90\_nslice3\_}\newline\texttt{fpw0.9\_nrw1.0v1.7\_10yrs},  \texttt{six\_stripe\_scale0.90\_nslice6\_}\newline\texttt{fpw0.9\_nrw0.0v1.7\_10yrs} & Tests of survey footprints with more emphasis on the Galactic plane, including rolling cadence in these variations in order to explore the impact on transients in the Galactic plane. The rolling cadence algorithm is different from the one in the above ``rolling'' simulations in the way it suppresses revisits within the same night and thus more effectively decrease the length of inter-night gaps between visits. \\
\hline 
\end{longtable*}

\subsection{Metrics Analysis Framework}
\label{sec:MetricsAnalysisFramework}

The LSST Metrics Analysis Framework \citep[MAF,][]{Jones2014} is a Python framework designed to easily evaluate aspects of the simulated survey strategies\footnote{\url{https://sims-maf.lsst.io/}}. MAF computes
simple quantities, such as the total co-added depth or number of visits, but it can also be used to construct more
complex metrics tailored towards specific science cases such as our new metric\footnote{\url{https://github.com/ninahernitschek/LSST_cadencenote}} \citep{Hernitschek2021}.
To facilitate metric analysis of the LSST OpSim outputs, NOIRLab's Community Science \& Data Centre (CSDC) provides the LSST MAF software packages and simulations outputs to their Astro Data Lab science platform. As this is the most straightforward way to access LSST MAF software packages and simulations outputs, we recommend doing so.
However, users without access to Astro Data Lab can download the simulations from \url{http://astro-lsst-01.astro.washington.edu:8081/} and run the metrics by following the installation instructions provided\footnote{\url{https://www.lsst.org/scientists/simulations/maf}}.

\section{Results: Survey Strategy Metric and Evaluation} 
\label{sec:Results}

In this section, we describe the metric we developed to evaluate the feasibility of recovering light-curve modulation in RR Lyrae stars, as well as the results from evaluating this metric on several OpSim FSB 1.7 databases.

\subsection{The Metric}
\label{sec:TheMetric}

Using the most recent survey simulations (from the LSST OpSim FSB 1.7 release), in this work we develop a metric to assess the effectiveness of different cadences for the LSST survey regarding detection of amplitude, period, phase modulation effects (e.g., Blazhko effect) in RR Lyrae stars. We here focus on the optimization of the WFD survey, so for this reason we don't look at simulations that deal with testing different cadences in the DDF and mini-surveys. All simulations, however, contain DDF and mini-surveys, see description of simulations in Tab. \ref{tab:simulationfamilies}.

To evaluate the feasibility of recovering light-curve modulation in RR Lyrae stars, such as caused by the Blazhko effect, we have developed a metric \texttt{PeriodicStarModulationMetric}. 
While our primary purpose is studying the detection of the Blazhko effect, this metric not solely aims at that purpose, but at evaluating variable star light curves from short time intervals in general and thus will also be helpful for the detection of other variable stars used as standard candles. We evaluate our metric at OpSim FSB 1.7 databases \texttt{baseline\_nexp2\_v1.7\_10yrs.db} (baseline), \texttt{rolling\_scale0.2\_nslice2\_v1.7\_10yrs.db} (rolling cadence) and \texttt{pair\_times\_55\_v1.7\_10yrs.db} (pair times cadence), see
the OpSim1.7 Summary Information Document and Tab. \ref{tab:simulationfamilies}.

The goal of our metric is to calculate the fraction of recovered RRab and RRc stars based on the light curve length (time interval within the first 2 years of LSST) and distance modulus. The metric is based on the \texttt{sims\_maf\_contrib} \texttt{PeriodicStarMetric} metric, which was modified in a way to reproduce attempts to identify a change in period, phase or amplitude in RR Lyrae stars.    
We have not implemented this modulation in the curve itself, as the modulation can take very different forms. Instead, we took a more general approach to investigate how well we can
identify period, phase or amplitude from a variable star's light curve on rather short baselines (15, 20, 30, 50 days). This attempt is also useful for other purposes, i.e. if we want to test whether we can just recover period, phase and amplitude from short baselines at all, without necessarily having in mind to look for light-curve modulations.

Like in \texttt{PeriodicStarMetric}, the light curve of an RR Lyrae star, or a periodic star in general, is approximated as a simple sin wave. (A future version might make use of light-curve templates to generate light curves, see e.g. the RRab and RRc light curves from \cite{Sesar2012}.)
Instead of evaluating the complete light curve at once, we split light curves into time intervals.
We then measure the fraction of light curves whose parameters are recovered correctly (within a given tolerance) from those time intervals.
Two other modifications we introduced for the \texttt{PeriodicStarModulationMetric} are:
In contrast to \texttt{PeriodicStarMetric}, we allow for a random phase offset to mimic observation starting at random phase.
Also, we vary the periods and amplitudes within $\pm$10 \% to allow for a more realistic sample of variable stars.
The metric is calculated using HEALPix\footnote{\url{http://healpix.sourceforge.net/}} \citep{Gorski2005} maps, with pixel resolution of 7$^{\circ}$.33 (achieved using the HEALPix resolution parameter Nside = 8).

To test whether the parameters amplitude, period and phase offset of a simulated light curve can be correctly recovered, we run
a simple \texttt{curve\_fit} (\texttt{scipy.optimize}).

We evaluate our metric for different values of the distance modulus (17.0, 18.0, 19.0, 20.0, 21.0, 22.0) as well as different length of time intervals
(15, 20, 30, 50 days) on sin-wave light-curves with amplitudes and periods typical for RRab and RRc stars.

Code relevant for the metric \texttt{PeriodicStarModulationMetric} as well as the figures shown here can be found in our GitHub repository \citep{Hernitschek2021}.

\subsection{Analysis of Metric Results} 
\label{sec:AnalysisOfMetricResults}

\begin{figure*}
  \includegraphics[width = 1\textwidth]{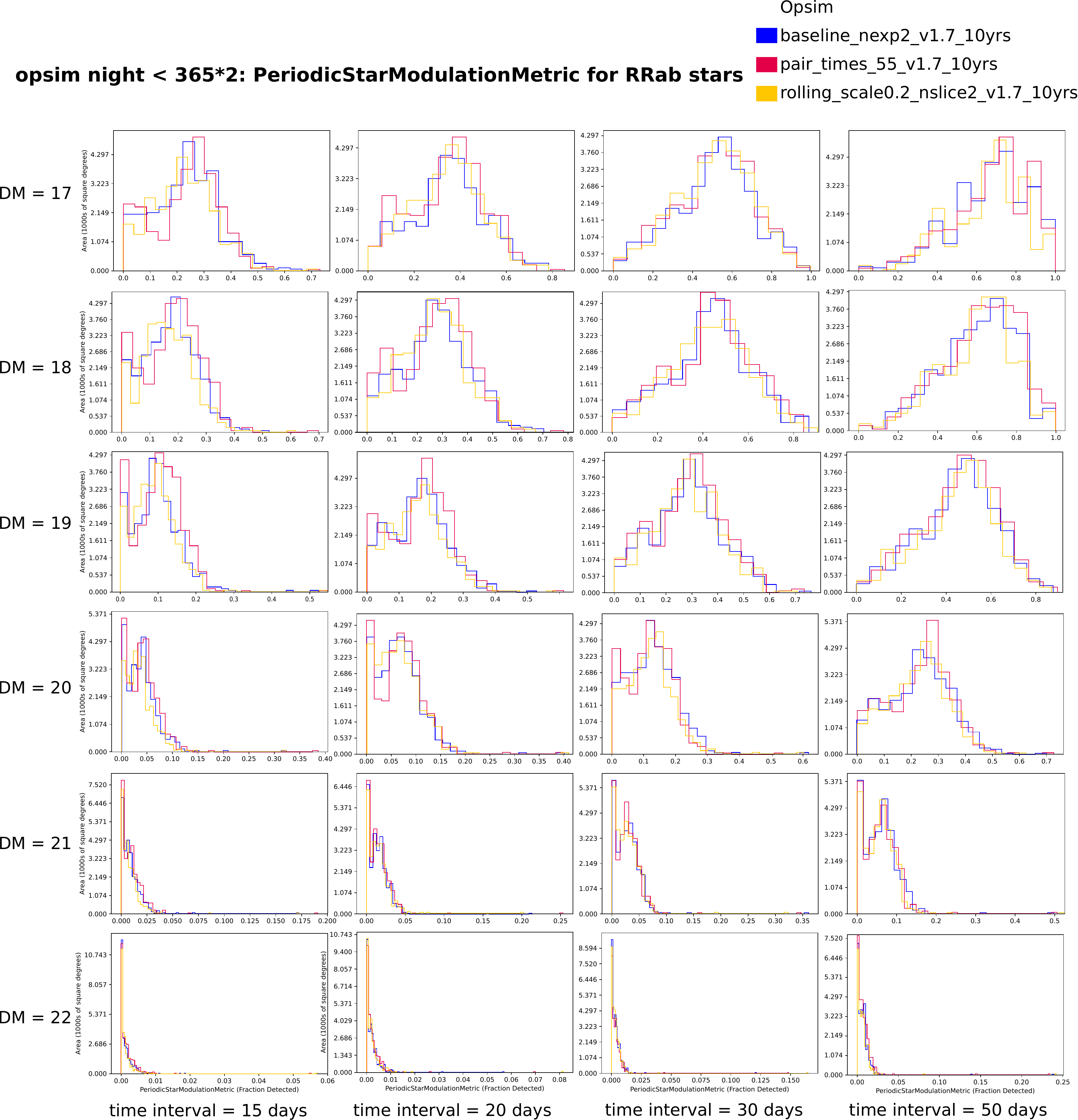}
  \caption{Comparison of metric evaluation for RRab light curves. We plot the area (in 1000s of square degrees) for which a given fraction of RRab stars with a given distance modulus (17 to 22) from a light curve with a given time interval (15 to 55 days) can be identified. The metric was evaluated on simulated 2-year light curves for OpSim 1.7 databases \texttt{baseline\_nexp2\_v1.7\_10yrs.db} (baseline, blue histograms), \texttt{pair\_times\_55\_v1.7\_10yrs.db} (pair times cadence, red histograms) and \texttt{rolling\_scale0.2\_nslice2\_v1.7\_10yrs.db} (rolling cadence, orange histograms).}  
    \label{fig:RRab}
\end{figure*}

\begin{figure*}
  \includegraphics[width = 1\textwidth]{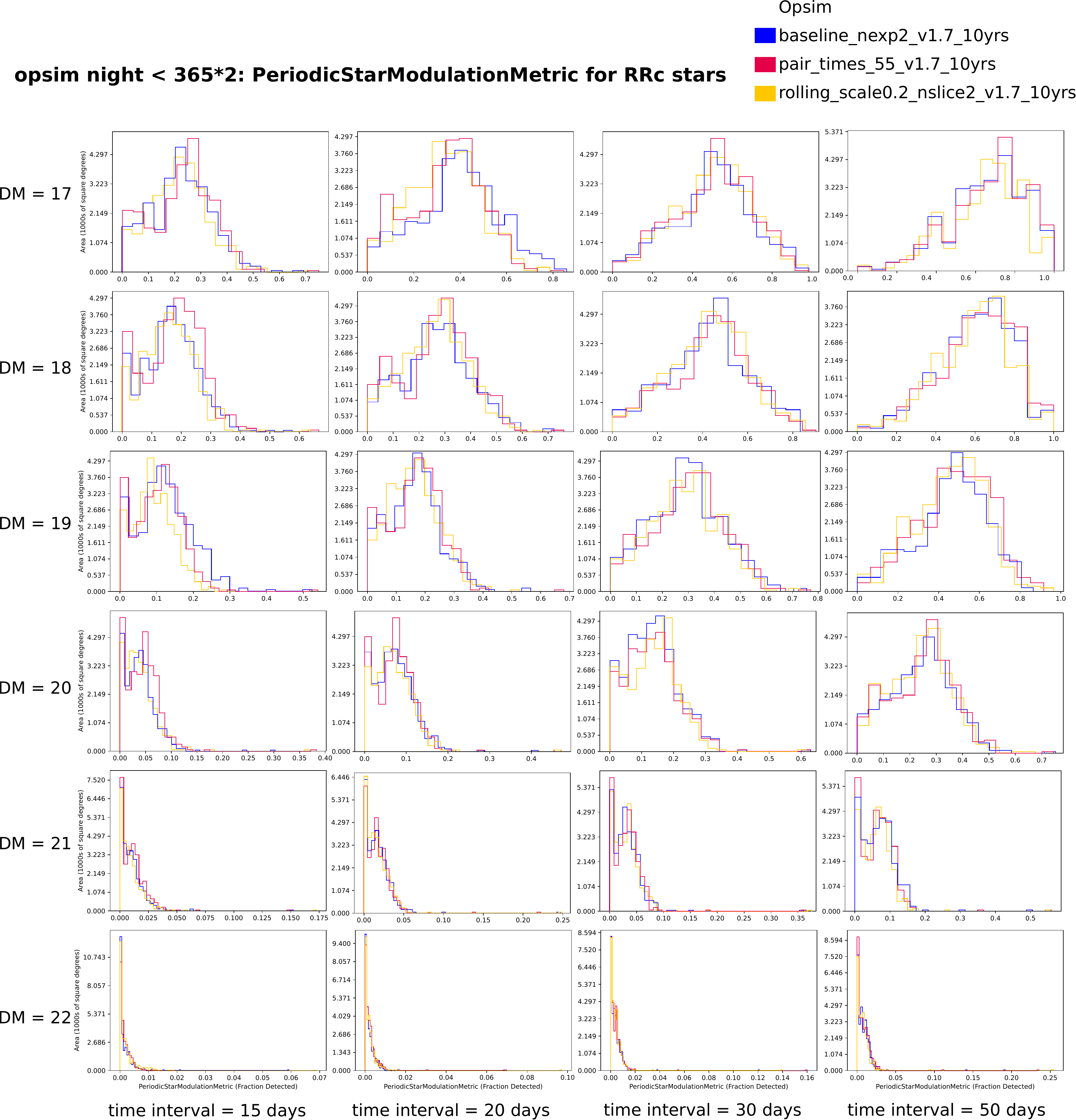}
  \caption{Comparison of metric evaluation for RRc light curves. We plot the area (in 1000s of square degrees) for which a given fraction of RRc stars with a given distance modulus (17 to 22) from a light curve with a given time interval (15 to 55 days) can be identified. The metric was evaluated on simulated 2-year light curves for OpSim 1.7 databases \texttt{baseline\_nexp2\_v1.7\_10yrs.db} (baseline, blue histograms), \texttt{pair\_times\_55\_v1.7\_10yrs.db} (pair times cadence, red histograms) and \texttt{rolling\_scale0.2\_nslice2\_v1.7\_10yrs.db} (rolling cadence, orange histograms).}  
    \label{fig:RRc}
\end{figure*}

We evaluate our metric on simulated 2-year light curves for OpSim FSB 1.7 databases \texttt{baseline\_nexp2\_v1.7\_10yrs.db} (baseline), \texttt{rolling\_scale0.2\_nslice2\_v1.7\_10yrs.db} (rolling cadence) and \texttt{pair\_times\_55\_v1.7\_10yrs.db} (pair times cadence).
Light curves were simulated both for RRab and RRc stars. Plots showing an evaluation of the metric for RRab and RRc stars can be found in Fig.~\ref{fig:RRab} and Fig.~\ref{fig:RRc}, respectively. 

In both Figures \ref{fig:RRab} and \ref{fig:RRc}, we plot the area which a given fraction of stars with a given distance modulus (17 to 22) from a light curve with a given time interval (15 to 55 days) can be identified. As we focus on early science with LSST, the metric was evaluated on simulated 2-year light curves for OpSim 1.7 databases \texttt{baseline\_nexp2\_v1.7\_10yrs.db} (baseline) which always should be used for comparison, and two observing strategies which seem the most promising regarding variable star science, \texttt{rolling\_scale0.2\_nslice2\_v1.7\_10yrs.db} (rolling cadence) and \texttt{pair\_times\_55\_v1.7\_10yrs.db} (pair times cadence).

\begin{figure*}
\begin{center}
  \includegraphics[width = 0.7\textwidth]{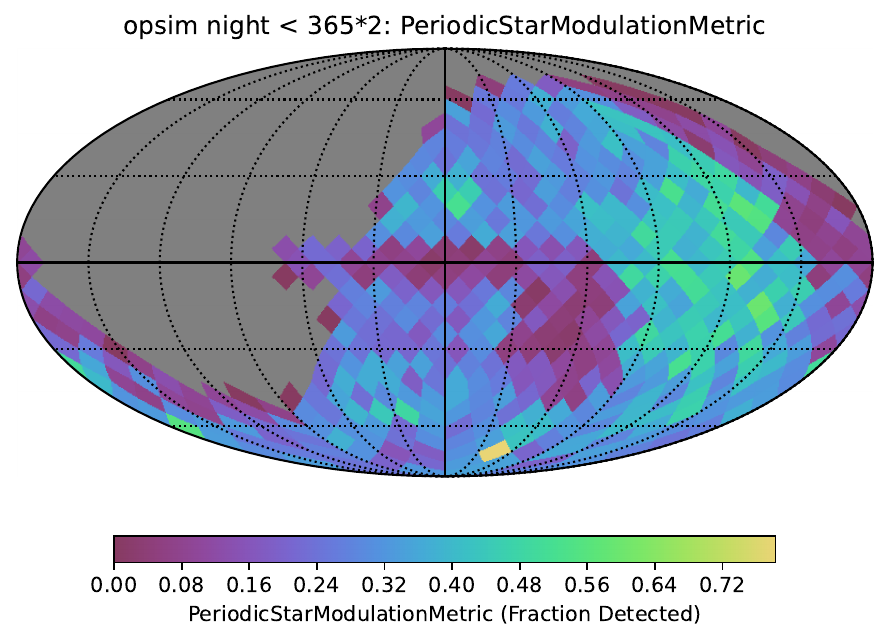}
\end{center}
  \caption{To illustrate how the histograms in Fig.~\ref{fig:RRab} and Fig.~\ref{fig:RRc} were calculated, this Figure shows a single metric result for a distance modulus of 18.0, and a time interval of 20 days for RRab-like light curves using the pair times cadence. This corresponds to the red histogram in the second panel of the second row in Fig.~\ref{fig:RRab}. The sky map is plotted in Galactic coordinates. Each HEALPix cell covers 7$^{\circ}$.33 (achieved using the HEALPix resolution parameter Nside = 8).
We higlight that the fraction of correctly recovered period, phase and amplitude (here as ``Fraction Detected'') varies over the survey footprint due to coverage and light curve length. The generation of sky maps as well as histograms is handled automatically within the 
\texttt{lsst.sims.maf.metricBundles} Python package provided by the MAF. For further reference, we recommend the documentation described in Sec. \ref{sec:AnalysisOfMetricResults}.}  
    \label{fig:RRab_18mag_20days_skymap}
\end{figure*}

To illustrate how the histograms in those Figures were calculated, 
Fig.~\ref{fig:RRab_18mag_20days_skymap} shows a single metric result for a distance modulus of 18.0, and a time interval of 20 days for RRab-like light curves using the pair times cadence. This corresponds to the red histogram in the second panel of the second row in Fig.~\ref{fig:RRab}.

\clearpage

\subsubsection{LSST Baseline Survey Results} 
\label{sec:LSSTBaselineSurveyResults}

In Fig. \ref{fig:RRab}, we plot the area (in 1000s of square degrees) for which a given fraction of RRab stars can be correctly detected, i.e. the light curve parameters retrieved within the allowed tolerances. We do this for time intervals of 15 to 50 days, and the distance modulus spanning 17 to 22.

As expected, the area over which a relevant fraction of RRab and RRc stars can be correctly recovered from such short time intervals drops significantly for a distance modulus $>$ 20.
For a distance modulus up to 19, for more than half of the survey footprint more than half of the light curves can be fit correctly using time intervals of 30 days.
For a distance modulus of 21, we have to move to a time interval of 50 days to get a correct fit for 10 \%.

Also, as expected, we find an increase of the recovery rate over time interval length. Here we want to highlight that even for a distance modulus of 20, over half of the survey footprint more than 30 \% of the light curves can be recovered correctly from a time interval of 50 days. For a lower distance modulus, the same recovery rate can be achieved for time intervals from 15 - 20 days.
As the modulation due to the Blazhko effect happens on timescales from weeks to months, using a time interval larger than 20 days is reasonable.

As we are especially interested in RR Lyrae stars in the outer halo, we are dealing with a distance modulus $>$ 19.

We now take a look at the influence of different survey strategy choices.

\subsubsection{Alternative Survey Strategy} 
\label{sec:AlternativeSurveyStrategy}

For the rolling cadence (evaluated only on \texttt{rolling\_scale0.2\_nslice2\_v1.7\_10yrs.db}), we find a sometimes (depending on distance modulus and time interval) slightly higher, sometimes slightly lower fraction of recovered light curves per area than for the baseline survey, but always less recovered light curves per area than for the the pair times cadence (\texttt{pair\_times\_55\_v1.7\_10yrs.db}).

In this comparison, we thus see as slight advantage of the pair times cadence over the other cadences tested.

Varying the pair time changes the overall number of filter changes per night, so longer pair times (55 minutes in our case) result
in more visits overall in the survey. In addition, this matches better the typical time scale of RR Lyrae light curves: The standard baseline attempts pairs at 22 minutes, whereas here we have chosen 55 minutes as timeline over which RR Lyrae star vary significantly.

The survey strategy of a rolling cadence means that some parts of the sky receive a higher number of visits during an `on' season,
followed by a lower number of visits during an `off' season, while during the first as well as the last year and half, the sky is covered uniformly as normal. We have so far not tested the influence of cadence on a full 10-year survey, but assume from our tests so far that the influence of a rolling cadence might be worse.

\section{Discussion: Optimization of Filter and Cadence Allocation} 
\label{sec:OptimizationofFilterandCadenceAllocation}
In this section, we discuss our results drawn from the analysis of the metric on the simulated observing strategy, and highlight our recommendations for the observing strategy choice regarding our science goals.
 
As RR Lyrae stars are fairly blue stars (spectral types A and F), the bluer filters ($u$, $g$) are particularly important for RR Lyrae and we thus want to have a high cadence in those filters. 
Detecting and characterizing relatively faint RR Lyrae stars in the old Milky Way's halo would in addition benefit from an increased $u$ band exposure time of 1$\times$50 sec.

The relevant simulation runs are in the family `u\_long' which tests different $u$ band exposure times, and filter\_cadence which adds $g$ band exposures with the primary goal  to improve the discovery for longer timescale transients like supernovae (see Tab. \ref{tab:simulationfamilies}).
Therefore, we recommend that the number of $u$, $g$ observations is increased in the WFD cadence plan to benefit the transient and variable star science.

In addition, the simulation family `pair\_times' deals with the question of obtaining two visits in a pair in the same vs. in different filters.

Observations in different filters are helpful for classification based on colors, for example to first identify RR Lyrae stars (and other standard candles) when light curves are too sparse to calculate periods. However, our attempt of getting precise periods, phases and amplitudes will benefit from having more observations in the same filter.
The simulation with a pair-spacing cadence of 55 days (\texttt{pair\_times\_55\_v1.7\_10yrs.db}) improves the recovery rate of RR Lyrae periods, phases and amplitudes from short-time light curves, as there is no significant variability on e.g. the 20-minute baseline.
Likely visit pairs more widely spaced would improve the ability to recover those light-curve parameters.
 
We investigated whether our metric benefits from cadence allocations as present in the simulation family `rolling'.
A rolling cadence means that some parts of the sky receive a higher number of visits during an `on' season, followed by a lower number of visits during an `off' season.
As this scenario provides higher-cadence light curves, the rolling cadence could benefit variable stars investigation, especially in the Galactic Plane. However, for variability analysis of RR Lyrae stars, we are looking for high-latitude objects in the old Milky Way's halo.
Our simulations for the rolling cadence have shown that the results are worse than for the baseline survey strategy.

This also agrees with other more general analysis showing the coverage of one-day timescales (see Cadence Note Bellm et al.\footnote{\url{https://docushare.lsst.org/docushare/dsweb/Get/Document-37644/Delta_T_2021.pdf}}): They find that the \texttt{rolling\_scale\*} and \texttt{alt\_roll} simulations have very poor (sub-percent) coverage of one-day timescales, the \texttt{rolling\_nm\_scale1.0\_nslice2} result is close to the \texttt{baseline}, and 
\texttt{rolling\_nm\_scale0.90\_nslice3\_fpw0.9\_nrw1.0} approaches the \texttt{pair\_times\_55} simulation in its effective timescale coverage.\\

\section{Conclusions} 
\label{sec:Conclusions}

The LSST survey shows great potential for carrying out the science goals of mapping the Milky Way and exploring the transient and variable optical sky.
We here explored especially its possibilities to detect the characteristic modulations of amplitude, period and phase shown by many RR Lyrae stars, the
so-called Blazhko effect.
Systematically studying this effect, which is known since more than 100 years \citep{Blazhko1907, Kollath2011}, so far was difficult as all-sky surveys such as
PS1 3$\pi$ lack the necessary number of observations within the rather short time frame of this effect which ranges from weeks to months, whereas other surveys have the necessary cadence to clearly show the Blazhko effect, but have a relatively small footprint or highly selective input target lists, such as e.g. the TESS survey, making population studies difficult.

With the upcoming LSST survey, which will cover a wide area of the sky with deep, rapid observations, we see an exciting possibility in carrying out such population studies.

In this paper, we have developed a metric to investigate the impact of observing strategy choices on the detectability of the Blazhko effect. 

We compared the metric on simulation runs form the `pair\_times' and `rolling' families with those from the baseline observing strategy.
From our results, our first recommendation for the observing strategy choice regarding our (and likely similar) science goals is to have a higher number of subsequent observations in the same filter with a medium-length spacing, such as 55 days. This improves the recovery rate of RR Lyrae periods, phases and amplitudes in contrast to the baseline strategy, as there is no significant variability on e.g. the 20-minute baseline.\newline
Our second recommendation is to use no rolling cadence observing strategy, as this would only improve the light curve cadence for variable stars in the Galactic Plane, whereas RR Lyrae stars are high-latitude objects in the old Milky Way's halo. This also agrees well with similar investigations from other Cadence Notes submitted.

\begin{acknowledgements}

This paper was created in the nursery of the Rubin LSST Transient and 
Variable Star Science Collaboration 
\footnote{\url{https://lsst-tvssc.github.io/}}. The authors acknowledge the 
support of the Vera C. Rubin Legacy Survey of Space and Time Transient and 
Variable Stars Science Collaboration that provided opportunities for 
collaboration and exchange of ideas and knowledge. The authors are thankful for 
the support provided by the Vera C. Rubin Observatory MAF team in the creation 
and implementation of MAFs. The authors acknowledge the support of the LSST 
Corporations that enabled the organization of many workshops and hackathons 
throughout the cadence optimization process.

This research uses services or data provided by the Astro
Data Lab at NSF’s National Optical-Infrared Astronomy Research Laboratory. NOIRLab is operated by the
Association of Universities for Research in Astronomy
(AURA), Inc. under a cooperative agreement with the
National Science Foundation.

Software: LSST metrics analysis framework \citep[MAF][]{Jones2014}; Astropy (Astropy Collaboration et al.
2013); JupyterHub\footnote{\url{https://jupyterhub.readthedocs.io/en/stable/index.html}}.

\end{acknowledgements}

%% This command is needed to show the entire author+affiliation list when
%% the collaboration and author truncation commands are used.  It has to
%% go at the end of the manuscript.
%\allauthors

%% Include this line if you are using the \added, \replaced, \deleted
%% commands to see a summary list of all changes at the end of the article.
%\listofchanges

\end{document}